\documentclass{CVIS}

\usepackage{amsmath}

\begin{document}

\title{Deep Quality: A Deep No-reference Quality Assessment System}

\author{
\begin{tabularx}{\textwidth}{X X}
Prajna Paramita Dash & University of Waterloo, ON, Canada \\
Akshaya Mishra & Miovision Technologies Inc. \\
Alexander Wong & University of Waterloo, ON, Canada \\
\end{tabularx}
}

\maketitle

\begin{abstract}
Image quality assessment (IQA) continues to garner great interest in the research community, particularly given the tremendous rise in consumer video capture and streaming.  Despite significant research effort in IQA in the past few decades, the area of no-reference image quality assessment remains a great challenge and is largely unsolved.  In this paper, we propose a novel no-reference image quality assessment system called Deep Quality, which leverages the power of deep learning to model the complex relationship between visual content and the perceived quality.  Deep Quality consists of a novel multi-scale deep convolutional neural network, trained to learn to assess image quality based on training samples consisting of different distortions and degradations such as blur, Gaussian noise, and compression artifacts.  Preliminary results using the CSIQ benchmark image quality dataset showed that Deep Quality was able to achieve strong quality prediction performance (\textbf{89\%} patch-level  and \textbf{98\%} image-level prediction accuracy), being able to achieve similar performance as full-reference IQA methods.
\end{abstract}

\section{Introduction}

The advances in multimedia technologies have revolutionized the way we capture, process, transmit and store digital media such as images and videos.  Using smart multi-media devices, ranging from digital cameras to smartphones, people capture and share billions of images and videos each day.  This tremendous rise in the volume of digital images and videos being captured has led to an ever-increasing demand for ways to assess the quality of the captured images and videos, particularly in the broadcast industry where video data must undergo different transcoding and compression processes to manage data bandwidth while maintaining a certain level of quality in a consistent manner to meet consumer demands.  As a result, \textit{image quality assessment (IQA)} has garnered great interest in the research community~\cite{Wang1,Mittal}.  IQA methods can be generally grouped into three main categories: 1) full-reference~\cite{George,Claudio,Shao,Lee}, 2) reduced reference~\cite{Wang2}, and 3) no-reference~\cite{Mittal,Gu}.  While tremendous progress has been made in the area of full-reference and reduced reference IQA methods, such methods are not suitable for scenarios where there is no reference information to leverage.  As such, despite significant research effort in IQA in the past few decades, the area of no-reference image quality assessment remains a great challenge and is largely unsolved.
\begin{figure}[h!]
\begin{center}
\includegraphics[scale = 0.2]{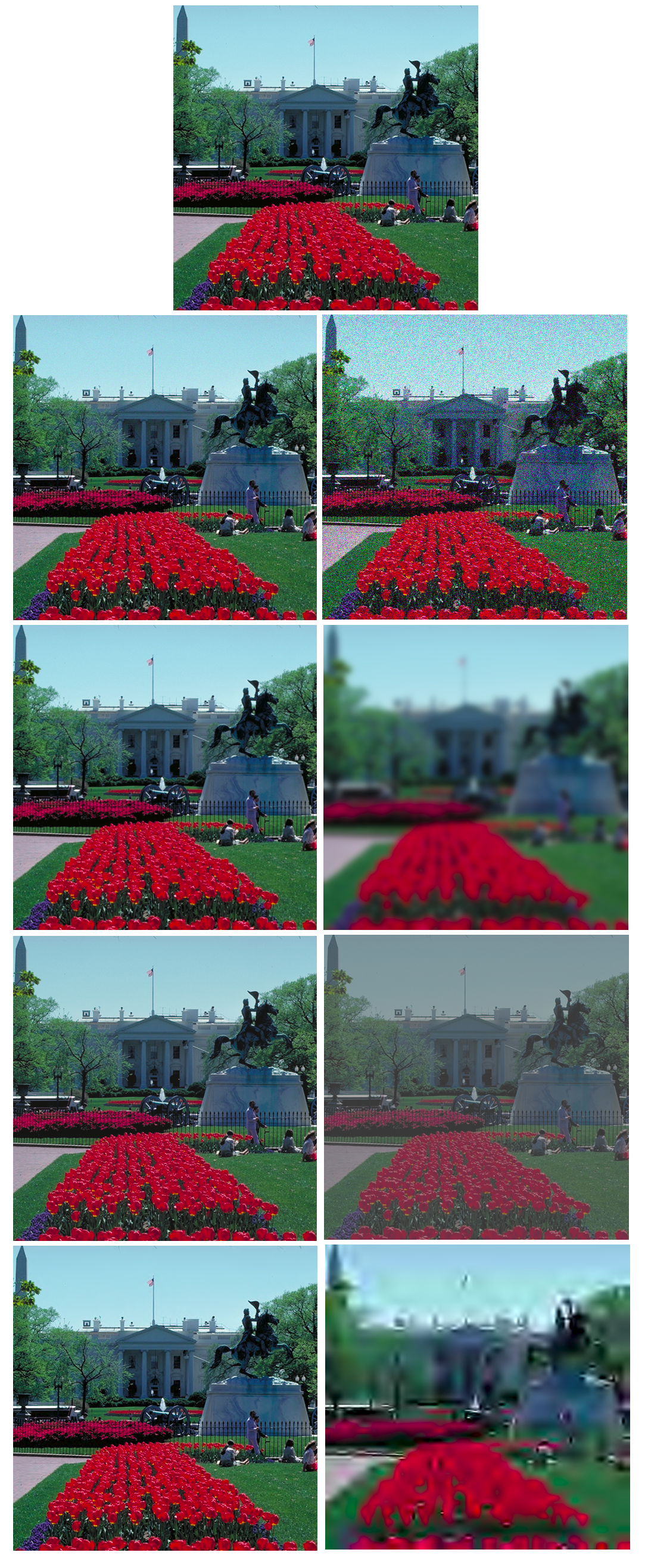}
\end{center}
\caption{Demonstration of several degradation models on an example image from the CSIQ dataset~\cite{Larson}. From top to bottom row: i) additive Gaussian pink noise, ii) Gaussian blur, iii) global contrast, and iv) JPEG-2000 compression. First and second columns show different degradation levels, where first column has lowest degradation ($c_0$) and right column has worst degradation ($c_4$) }
\label{fig1}
\end{figure}
Motivated by this challenge, we propose a novel no-reference image quality assessment system called Deep Quality, which leverages the power of deep learning to model the complex relationship between visual content and the perceived quality.  Deep Quality attempts to learn this complex relationship based on the appearance of the images across different scales via a novel multi-scale deep convolutional neural network, trained to learn to assess image quality based on training samples consisting of different distortions and degradations such as blur, Gaussian noise, and compression artifacts.

\section{Methodology}
\begin{figure}[h]
\begin{center}
\includegraphics[scale = 1.5]{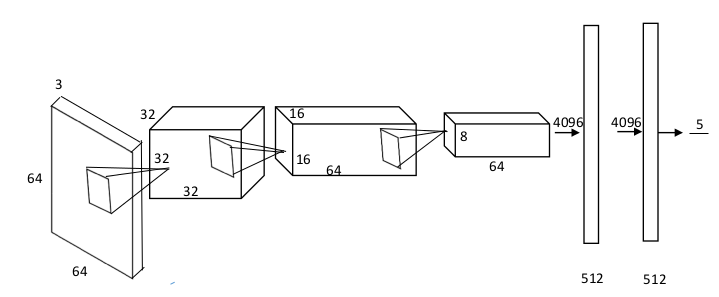}
\end{center}
\caption{Deep Quality's network architecture. The network have three convolutional, ReLU, and maxpool layers.  An input patch of size $64 \times 64$ is mapped to a five-class classification task, where each class corresponds to a particular image quality grade.}
\label{fig2}
\end{figure}
Deep Quality formulates the image quality assessment as a learning problem, where, for an image $I$ of size $n \times m$, the classifier is defined as
\begin{equation}
C(I) = \sum_{i=0}^{n\times  m}(w_i\times f_i(I(X(i))),
\end{equation} where $f_i$ is defined as a classifier for the image patch $I(X(i))$ at location $X(i)$ with weight $w_i$. Motivated by the classification performance of convolutional neural networks, Deep Quality uses a convolutional neural network approach to describe $f_i$. A back-propagation-based gradient descent algorithm can be employed to find the parameters $w_i$ and $f_i$ of classifier $C(I)$. An overview of Deep Quality can be described as follows.  First, an image is decomposed into several local patches. Second, an image patch quality deep neural network is used to classify individual image patches into five classes, each corresponding to a different image quality grade: $c_0\cdot. c_4$, with the best and worst grade are defined as $c_0$  and $c_4$. Finally,  Deep  Quality combines the scores of each patch classifier using a linear classifier to estimate the global image quality of an input image.

The network architecture of the Deep Quality deep convolutional neural networks is demonstrated in Fig.~\ref{fig2}.  The three important components of Deep Quality are described as local image patch pooling, local image patch quality classifier, global image quality estimator. The goal of local image patch pooling is to select $l$ number local image patches from $g$ number of images. To achieve this task, we first generate all possible patches using a sliding window approach, sort those patches based upon their variance, and then select $l$ lowest variance patches for local patch quality classification. The local image patch classifier is realized using 3 convolutional, 3 ReLU, 3 maxpool, and 2 fc layers. The three convolutional layers use $5 \times 5$, $3 \times 3$, and $3 \times 3$ kernels with a single stride, respectively. The three maxpool layers uses a window of size $[1, 2, 2, 1]$.

\section{Experimental setup and evaluations}
\subsection{Datasets}
The effiacy of the proposed Deep Quality system for no-reference IQA is investigated using the CSIQ image quality benchmark dataset \cite{Larson}.  The CSIQ dataset consists of 30 different types of natural reference images, with each reference image degraded using five different distortion types: i) JPEG compression, ii) JPEG-2000 compression, iii) global contrast decrements, iv) additive pink Gaussian noise, and v) Gaussian blurring. Some of those distorted images are shown in Fig.~\ref{fig1}. Each distortion type was applied at four to five different levels of distortion, resulting in 866 different distorted images.  The CSIQ dataset~\cite{Larson} also consists of a corresponding set of 5000 subjective ratings from 35 different observers reported in the form of DMOS.  For the proposed Deep Quality system, the DMOS scores are mapped into 5 different levels of image quality for evaluation purposes.

\subsection{Experimental Results and Discussions}
For learning local patch level image quality, Deep Quality have sampled $60,000$ image patches from the $866$ distorted input images using a variance pooling mechanism. The variance pooler uses a overlapping sliding window of size $64\times 64$ to generate a large set of local patches and selects $70$ low variance local patches from each images, generating $60,000$ total image patches. These $60,000$ image patches are splitted into two sets: i) training set amounting $50,000$ training samples, and ii) $10,000$ testing samples. For training, we have used sparse cross entropy and logistic function as data cost and have used an $l$2 norm of the weights of two fc layers as the regularization cost. The network is trained for $100$ epochs. The training local patch-level training accuracy and testing accuracy after 100 epochs are \textbf{95.5\%} and \textbf{89\%}, respectively. Furthermore, when the score of each patch of an input image is combined with a linear classifier, Deep Quality's image-level accuracy exceeds \textbf{98\%} which is comparable to state-of-the-art full-reference image quality algorithms.

In our experiments, we have trained and tested our model for each noise type separately as well as all noise types combined.  We observed that individual noise types can be trained with simpler networks, while to train the system to handle combined noise types required a more complex network. We also observed that region pooling has a high impact on image patch classification accuracy. Variance-based pooling worked best for additive pink Gaussian but did not work well for blur and JPEG-type distortions.

\section{Conclusion and Future Work}
In this paper, We have developed and implemented a deep non-reference image quality assessment system called Deep Quality. Preliminary results show that Deep Quality was able to achieve \textbf{89\%} patch-level  and \textbf{98\%} image-level  accuracy using the CSIQ~\cite{Larson} dataset.

Our preliminary implementation of Deep Quality decoupled the terms of the Deep Quality estimator and have optimized each step separately. Simultaneous optimization of these terms in a deep learning framework could provide a better quality estimator and is left as future work.

\section{Acknowledgment}
This work was supported by the Natural Sciences and Engineering Research Council (NSERC) of Canada, and in part by the Canada Research Chairs program.

\end{document}